\documentstyle[twocolumn,aps,psfig]{revtex}

\begin{document}
\draft
\flushbottom
\twocolumn[
\hsize\textwidth\columnwidth\hsize\csname @twocolumnfalse\endcsname

\title{ Cylindrical surface plasmon mode coupling as a three-dimensional Kaluza-
Klein theory}
\author{Igor I. Smolyaninov }
\address{ Department of Electrical and Computer Engineering \\
University of Maryland, College Park,\\
MD 20742}
\date{\today}
\maketitle
\tightenlines
\widetext
\advance\leftskip by 57pt
\advance\rightskip by 57pt

\begin{abstract}
Cylindrical surface plasmon (CSP) mode coupling in a nanowire or a nanochannel 
is described in terms of three-dimensional Kaluza-Klein theory in which the 
compactified third dimension is considered to be the angular coordinate of the 
metal cylinder. Higher ($n\neq 0$) CSP modes are shown to posses a quantized 
effective charge proportional to their angular momentum $n$. In a 
nanowire these slow moving charges exhibit strong long-range interaction via 
fast massless CSPs with zero angular momentum ($n=0$). Such a mode-coupling 
theory may be used in description of nonlinear optics of cylindrical nanowires 
and nanochannels (for example, in single-photon tunneling effect), and may be 
further extended to describe interaction of electrons with nonzero angular momenta in thin gold wires and carbon nanotubes.   



\end{abstract}

]

\narrowtext

\tightenlines

Quantum communication and information processing require unusual sources of 
light with strong quantum correlations between single photons. This necessity 
drives strong current interest in exploration and understanding of nonlinear 
optical phenomena at single photon levels. One example of such phenomena is 
photon blockade in a nonlinear optical cavity \cite{1} introduced in close 
analogy with the well-known phenomenon of Coulomb blockade for quantum-well 
electrons. Another recent example is the proposed observation of sound waves 
and, possibly, superfluid state in a weakly interacting "photon gas" in a 
nonlinear Fabry-Perot cavity \cite{2}. In the latter proposal photons in the 
nonlinear cavity are treated as massive particles weakly interacting with each 
other via some unspecified nonlinear optical mechanism.

Very recently strong evidences of nonlinear optical effects such as single-
photon tunneling \cite{3} and light-controlled photon tunneling \cite{4} 
occurring at very low light intensities in natural nanometer-scale pinholes  and 
artificial cylindrical nanochannels drilled using focused ion beam milling in 
thick gold films and membranes, and filled with nonlinear polymer have been 
observed. Transmission of nanoholes in gold and silver films is commonly 
accepted to happen via excitation of intermediary surface plasmons (SP) 
propagating along the metal film surfaces, and cylindrical surface plasmons 
(CSP) excited in the cylindrical nanochannels \cite{4}. In the case of long 
cylindrical nanochannels filled with nonlinear material theoretical description 
of nonlinear optical transmission of the nanochannels requires understanding of 
CSP behavior (such as mode propagation and interaction) at a level of single 
optical quanta in the presence of strong nonlinear optical interaction between 
them. So far to my knowledge, no such theory has been developed. 

In this Letter I analyze CSP behavior in two closely related cylindrical 
geometries, such as a long cylindrical nanochannel drilled in a metal that 
supports surface plasmons (such as gold and silver) and filled with nonlinear 
material, and a long cylindrical metal nanowire embedded in a nonlinear 
material. In both geometries a specific case of chiral nonlinear media will be 
addressed (it is important to mention that metal itself becomes a chiral 
material in an external magnetic field \cite{5}, and all metal interfaces 
exhibit pronounced nonlinear behavior, so that chirality and nonlinear optical 
effects must be considered in cylindrical surface plasmon propagation). Since 
CSPs may be considered as if they live in a "three-dimensional space-time" with 
two spatial dimensions being a very long z-dimension along the axis of the 
cylinder, and a small "compactified" $\phi $-dimension along the circumference 
of the cylinder, the theory of CSP mode propagation and interaction is 
formulated similar to the three-dimensional Kaluza-Klein theory \cite{6}. In the 
most interesting cylindrical nanowire case, higher ($n>0$) CSP modes are shown 
to posses a quantized effective charge (not of electric origin) proportional to 
their angular momentum. These slow moving effective charges exhibit strong long-
range interaction via fast uncharged massless CSPs with zero angular momentum 
($n=0$). The character of CSP mode interaction in a nanochannel is somewhat 
similar, but in this case the zero angular momentum modes acquire mass, so that 
interaction of higher mode quanta becomes short-ranged.    
Such a mode-coupling theory may be further extended to describe mode propagation 
and interaction in other chiral optical waveguides, as well as interaction of 
nonzero angular momentum electron states in thin gold wires and carbon 
nanotubes.   

Let us recall the basic features of Kaluza-Klein theories.
In modern Kaluza-Klein theories \cite{6} the extra N-4 space-time dimensions are 
considered to be compact and small (with characteristic size on the order of the 
Planck length). The symmetries of this internal space are chosen to be the gauge 
symmetries of some gauge theory \cite{7}, so a unified theory would contain 
gravity together with the other observed fields. In the original form of the 
theory a five-dimensional space-time was introduced where the four dimensions 
$x^1, ..., x^4$ were identified with the observed space-time. The associated 10 
components of the metric tensor $g_{\alpha \beta }$ were used to describe 
gravity. After a compactified fifth dimension $x^5$ with a small circumference 
$L$ was added, the extra four metric components $g_{\alpha 5}$ connecting $x^5$ 
to $x^1, ..., x^4$ gave four extra degrees of freedom, which were interpreted as 
the electromagnetic potential (here we use the following convention for greek 
and latin indices: $\alpha = 1, ..., 4$; $i = 1, ..., 5$). An additional scalar 
field $g_{55}$ or dilaton may be either set to a constant, or allowed to vary. 

When a quantum field $\psi $ coupled to this metric via an equation
\begin{equation} 
\Box_5 \psi +a\psi =0
\end{equation}
is considered, where $\Box_5$ is the covariant five-dimensional d'Alembert 
operator, the solutions for the field $\psi $ must be periodic in the $x^5$ 
coordinate. This leads to the appearance of an infinite "tower" of solutions 
with quantized $x^5$-component of the momentum: 
\begin{equation}
q^5_n=2\pi n/L 
\end{equation}
where $n$ is an integer. In our four-dimensional space-time on a large scale such solutions with $n\neq 0$ interact with the electromagnetic potential $g_{\alpha 5}$ as charged particles with an electric charge $e_n$ and mass $m_n$:
\begin{equation}
e_n = \hbar q_n(16\pi G)^{1/2}/c
\end{equation}
\begin{equation}
m_n = \hbar (q_n^2-a)^{1/2}/c
\end{equation}
where $G$ is the gravitational constant (see for example the derivation in 
\cite{8}). For the purposes of discussion below let us follow this derivation 
when an angular coordinate $\phi ^5$ varying within an interval from 0 to 
$2\pi $ is introduced, so that 
\begin{equation}
\phi ^5 = 2\pi \frac{x^5}{L}   
\end{equation}

Now the metric can be written as
\begin{equation}
ds^2=g_{\alpha \beta }dx^{\alpha }dx^{\beta }+2g_{\alpha 5}dx^{\alpha }d\phi 
^5+g_{55}d\phi ^5d\phi ^5,
\end{equation} 
Here we are not interested in possible spatial dependence of $g_{55}$ and 
consider all $g_{i5}$ components to be independent of $\phi ^5$. 
Equation (1) with $a=0$ for the quantum field $\psi $ in this metric should be 
written as
\begin{eqnarray}
\frac{\partial }{\partial x^{\alpha }}(g^{\alpha \beta }\frac{\partial \psi 
}{\partial x^{\beta }})+\frac{\partial }{\partial x^{\alpha }}(g^{\alpha 
5}\frac{\partial \psi }{\partial \phi ^5})+ \nonumber \\
\frac{\partial }{\partial \phi ^5}(g^{5\alpha }\frac{\partial \psi }{\partial 
x^{\alpha }})+\frac{\partial }{\partial \phi ^5}(g^{55}\frac{\partial \psi 
}{\partial \phi ^5})=0
\end{eqnarray}
Since we assume that the $g_{i5}$ do not depend explicitly on $\phi ^5$, we 
should not address the terms like $\partial g^{i5}/\partial \phi ^5$. Thus, we 
may search for the solutions in the usual form as
$\psi = \Psi (x^{\alpha })e^{iq\phi ^5}$, where periodicity in $\phi ^5$ 
requires $q_n=n$. As a result, we obtain
\begin{equation}
\Box \psi -q_n^2\frac{1-g_{\alpha 5}g^{\alpha 5}}{g_{55}}\psi+2iq_ng^{\alpha 
5}\frac{\partial \psi}{\partial x^{\alpha }}+iq_n \frac{\partial g^{\alpha 
5}}{\partial x^{\alpha }}\psi = 0
\end{equation} 
This is the same as the Klein-Gordon equation in the presence of an 
electromagnetic field: in four-dimensional space-time it describes a particle of 
mass
\begin{equation}
m=\frac{2\pi \hbar q_n}{cg_{55}^{1/2}(q_n)},
\end{equation}
which interacts with a vector field $g^{\alpha 5}$ through a quantized charge 
$e_n \sim q_n$. In this theory the conservation of the electric charge is a 
simple consequence of the conservation of the $x^5$-component of the momentum. 
Similar Kaluza-Klein theory may be formulated in a three-dimensional space-time, 
which has one compactified spatial dimension. Again, quantized $\phi $-component 
of the momentum will play a role of an effective charge, which interacts with a 
two-component vector field $g^{\alpha 3}$.  

It is well-known that Maxwell equations in a general curved space-time 
background $g_{ik}(x,t)$ are equivalent to the macroscopic Maxwell equations in 
the presence of matter background with some nontrivial electric and magnetic 
permeability tensors $\epsilon _{ik}(x,t)$ and $\mu _{ik}(x,t)$ \cite{9}. Thus, 
strong similarity between the results of the three-dimensional Kaluza-Klein 
theory and the solutions of Maxwell equations in some quasi-three-dimensional 
cylindrical waveguide geometries may be expected: in both cases we consider a 
massless field in a geometry which has a compactified cyclic $\phi $-dimension 
in the presence of nontrivial space-time curvature or permeability tensors, 
respectively. This similarity is expected to be especially strong in the case of 
surface plasmon waveguides, since surface plasmons live in (almost) three-
dimensional space-times on the metal interfaces.  

Let us first consider general properties of Maxwell equations in a medium 
(media), which has cylindrically symmetric geometry. The insight gained from the 
previous discussion tells us that in order to look similar to the three-
dimensional Kaluza-Klein theory, the medium should discriminate between left- 
and right- circular polarized waves (the waves which have opposite angular 
momenta, and thus expected to posses opposite effective Kaluza-Klein charges). 
This means that the medium should be chiral or optically active. There are 
different ways of introducing optical activity (gyration) tensor in the 
macroscopic Maxwell equations. It can be introduced in a symmetric form, which 
is sometimes called Condon relations \cite{10}:
\begin{equation}  
\vec{D}=\epsilon \vec{E}+\gamma\frac{\partial \vec{B}}{\partial t}
\end{equation}
\begin{equation}  
\vec{H}=\mu ^{-1}\vec{B}+\gamma\frac{\partial \vec{E}}{\partial t}
\end{equation}
Or it can be introduced only in an equation for $\vec{D}$ (see \cite{5,11}). In 
our consideration I will follow Landau and Lifshitz \cite{5}, and for simplicity 
use only the following equation valid in isotropic or cubic-symmetry materials:
\begin{equation}  
\vec{D}=\epsilon \vec{E}+i\vec{E}\times \vec{g} ,
\end{equation}
where $\vec{g}$ is called the gyration vector. If the medium exhibits magneto-
optical effect and does not exhibit natural optical activity $\vec{g}$ is 
proportional to the magnetic field $\vec{H}$:
\begin{equation}  
\vec{g}=f\vec{H} ,
\end{equation}
where the constant $f$ may be either positive or negative. For metals in the 
Drude model at $\omega >>eH/mc$
\begin{equation}  
f(\omega )= -\frac{4\pi Ne^3}{cm^2\omega ^3}=-\frac{e\omega _p^2}{mc\omega ^3} ,
\end{equation}
where $\omega _p$ is the plasma frequency and $m$ is the electron mass \cite{5}.

Initially, let us consider a medium with $\vec{g}=\vec{g}(r,z,t)$ directed along 
the $\phi $-coordinate. Such a distribution may be produced, for example, in a 
medium exhibiting magneto-optical effect around a cylindrical nanowire if a 
current is passed through the wire. After simple calculations we obtain a wave 
equation in the form:
\begin{equation}  
\vec{\nabla }\times \vec{\nabla }\times \vec{B}=-\Delta \vec{B}=-\frac{\epsilon 
}{c^2}\frac{\partial ^2\vec{B}}{\partial t^2}+\frac{i}{c}\frac{\partial 
(\vec{\nabla }\times [\vec{E}\times\vec{g}])}{\partial t}  
\end{equation}
The z-component of this wave equation for a solution $\sim e^{in\phi }$ is:
\begin{eqnarray}
\frac{1}{r}\frac{\partial }{\partial r}(r\frac{\partial B_z}{\partial 
r})+\frac{\partial ^2B_z}{\partial z^2}-\frac{\epsilon \partial 
^2B_z}{c^2\partial t^2}-\frac{n^2}{r^2}B_z+ \nonumber \\
i\frac{ng}{rc}\frac {\partial (iE_z)}{\partial t}+\frac{in}{cr}(\frac{\partial 
g}{\partial t})(iE_z)=0
\end{eqnarray}
 
It can be re-written in the form similar to equations (1) and (8) as follows:
\begin{equation}
\hat{a_r} B_z+\Box _2B_z-\frac{n^2}{g_{\phi \phi 
}}B_z+in(\frac{g}{r})(\frac{\partial iE_z}{c\partial t})+in\frac{\partial 
(g/r)}{c\partial t}E_z=0,
\end{equation}
where $\hat{a_r}$ plays the role of factor $a$ in equation (1), and $g_{\phi \phi }=r^2$. Similarity of equations (8) and (17) becomes more evident if we 
identify the Kaluza-Klein vector field $g^{0\phi }$ as $g/r$, disregard terms 
higher than linear in $g$, and recall that for cylindrical surface plasmon modes 
$iE_z = \alpha B_z$, where the coefficient of proportionality $\alpha $ is real 
and is determined by the boundary conditions \cite{12}.
The missing factor of 2 in front of the fourth term in equation (17) originates 
from the asymmetric way of introducing the gyration in the macroscopic Maxwell 
equations (see equations (10-12) and the relevant discussion). Thus, from the 
macroscopic Maxwell equations we arrive to a picture of $n>0$ waveguide modes 
interacting with the "gyration potential" $g/r$ via quantized effective "chiral 
charges" proportional to the angular momenta of the modes. 

The dispersion law and electromagnetic field distribution of cylindrical surface 
plasmon modes of a cylindrical metal wire may be found in \cite{12}. The $\omega 
(k)$ of the $n=0$ CSP mode goes down to $\omega =0$ approaching the light line 
$\omega =kc/\epsilon ^{1/2}$ from the right, as $k\rightarrow 0$. The field of 
this mode has only the following nonzero components: $E_r$, $E_z$, and $H_{\phi 
}$, thus satisfying our initial requirements for the direction of $\vec{g}$, and 
if we recall equation (13), we notice that in the media exhibiting magneto-
optical effect, such as the cylindrical metal wire itself, the higher $n>0$ CSP 
modes interact with the $H_{\phi }$ field of the $n=0$ CSPs via their effective 
"chiral charges" proportional to $n$. Thus, CSP quanta with $n=0$ may be 
considered as massless (in the $k\rightarrow 0$ limit) quanta of the "gyration 
field". The dispersion laws of the CSP modes with $n>0$ start at some nonzero 
frequencies and intersect the light line (become nonradiative and de-couple from 
free-space photons) at some finite $\omega $. The modes of different $n$ are 
well separated from each other, and there is no crossing. In the $k\rightarrow 
\infty $ limit the $\omega (k)$ of all the CSP modes saturates at $\omega 
=\omega _p/2^{1/2}$. Thus, the group velocity $d\omega /dk$ of the higher modes 
goes to 0 in this limit: the "charged" quanta are slow.  

Let us now derive an analog of the Poisson equation for the "gyration potential" 
and "chiral charges" around the cylindrical metal wire. Let us search for the 
solutions of the wave equation (15) in the form $\vec{B}=\vec{B_0}+\vec{B_n}$ 
and $\vec{E}=\vec{E_0}+\vec{E_n}$, where $\vec{B_0}$ and $\vec{B_n}$ are the CSP 
fields with zero and nonzero ($n$) angular momenta, respectively, and the 
"gyration field" in the magneto-active media is obtained in a self-consistent 
manner as $\vec{g}=f(\vec{H}+\vec{B_0}+\vec{B_n})$, where $\vec{H}$ is a 
constant external field. We are interested in the solution for the field 
$\vec{B_0}$ in the limit $\omega _0 \rightarrow 0$ in the presence of the 
$\vec{B_n}$ field. The resulting nonlinear Maxwell equation may be simplified 
assuming that the field $\vec{B_n}$ is supposed to be the solution of linear 
Maxwell equation, and the terms proportional to $f^2$ and higher may be 
neglected. As a result, we obtain:
\begin{eqnarray}  
\Delta \vec{B_0}=-\frac{if}{c}\frac{\partial (\vec{\nabla }\times 
[\vec{E_n}\times\vec{B_n}])}{\partial t}-\frac{if}{c}\frac{\partial (\vec{\nabla 
}\times [\vec{E_0}\times\vec{B_n}])}{\partial t}- \nonumber \\
\frac{if}{c}\frac{\partial (\vec{\nabla }\times 
[\vec{E_n}\times\vec{B_0}])}{\partial t}      
\end{eqnarray}

Since the fields $\vec{B_0}$ and $\vec{B_n}$ are not supposed to be coherent, 
their products disappear after time averaging, and we are left with the 
equivalent Poisson equation:
\begin{equation}  
\Delta \vec{B_0}=\frac{f\omega _n}{c}\vec{\nabla }\times 
[\vec{E_n}\times\vec{B_n}]=\frac{4\pi f\omega _n}{c^2}\vec{\nabla }\times 
\vec{S_n} ,
\end{equation}
where $\vec{S_n}$ is the Pointing vector of the CSP field with the nonzero ($n$) 
angular momentum. As an interesting consequence of this equation, let us note 
that in an isotropic medium 
\begin{equation}  
\vec{\nabla}\times (\vec{\nabla }\times \vec{B_0}-\frac{4\pi f\omega _n 
}{c^2}\vec{S_n})=0 
\end{equation}
This observation lets us to conclude what is the real physical nature of the 
"chiral charge" in the normal language of four-dimensional physics. Due to 
magneto-optical effect, cylindrical surface plasmons with nonzero angular 
momenta behave as current loops bound to the metal wire. Unlike dipole-dipole 
interaction of the current loops in free space, the interaction of CSPs is 
mitigated by the presence of cylindrical surface plasmons with zero angular 
momentum. Thus, interaction of current loops becomes quasi one-dimensional and 
very long-ranged.  

Using the same approximations as before, we can also derive an analog of the 
Gauss theorem for the "chiral charges". Let us consider a cylindrical volume $V$ 
around a cylindrical metal wire (see Fig.1), such that the side wall of the 
volume $V$ is located very far from the wire and the CSP fields are zero at this 
wall. 

\begin{figure}[tbp]
\centerline{
\psfig{figure=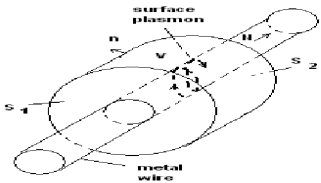,width=8.5cm,height=5.0cm,clip=}}
\caption{ Schematic view of a cylindrical metal wire, which supports cylindrical 
surface plasmon propagation. $\vec{N}$ is the chosen direction of the wire.
}
\label{fig1}
\end{figure}

Using the vector calculus theorem we can write
\begin{equation}
\int_{V}\vec{\nabla }\times \vec{S}d^3x=\int_{S}\vec{n}\times 
\vec{S}da=\int_{S_2}\vec{N}\times \vec{S}da-\int_{S_1}\vec{N}\times \vec{S}da,
\end{equation}
where $S$ is the closed two-dimensional cylindrical surface bounding $V$, with 
area element $da$ and unit outward normal $\vec{n}$ at $da$, $S_1$ and $S_2$ are 
the front and the back surfaces of $V$, and $\vec{N}$ is the chosen direction of 
the wire. Using equation (20) we obtain
\begin{eqnarray}
\int_{V}\frac{4\pi f\omega _n}{c^2}\vec{\nabla }\times 
\vec{S}d^3x=\int_{S_2}\vec{N}\times [\vec{\nabla }\times \vec{B_0}]da- \nonumber 
\\
\int_{S_1}\vec{N}\times [\vec{\nabla }\times \vec{B_0}]da
\end{eqnarray}
Since $\vec{N}\times [\vec{\nabla }\times \vec{B_0}]=\partial B_{0\phi }/\partial z$, we see that a "chiral charge" produces a local step in the 
"gyration field". If this effective Gauss theorem would be precise at any 
distance, we would come up with an unphysical result that a localized "chiral 
charge" would produce an infinite field $B_{\phi }$ at infinity. In reality, one 
has to take into account all the effects that we neglected while deriving the 
effective Poisson equation and Gauss theorem, such as higher than quadratic 
terms in the nonlinear Maxwell equations, and also take into account final 
lifetime of the cylindrical surface plasmons. Nevertheless, the underlying 
Kaluza-Klein mechanism of the "chiral charge" interaction lends credibility to 
the conclusion that CSPs of a cylindrical metal nanowire exhibit very strong 
long-range interaction. The case of cylindrical nanochannel is somewhat 
different, since in this case the zero angular momentum mode has nonzero 
frequency in the limit $k\rightarrow 0$ (in other words it acquires some 
effective mass), so that interaction of higher mode quanta via exchange of zero-
angular momentum CSPs becomes more short-ranged. Similar mode-coupling theory 
may be further developed to describe mode propagation and interaction in other 
chiral optical waveguides.   

Let us estimate the strength of chiral charges interaction. Using equations (17) 
and (22) within the approximations described above the potential energy of two 
equal chiral charges $n$ separated by distance $z$ can be written as
\begin{equation}
W=z\frac{2\pi \hbar \omega _nf^2nS_{n\phi }}{\epsilon rc},
\end{equation}
where $S_{n\phi }=n\omega _n(\epsilon E_z^2+B_z^2)/(4\pi \beta _n^2r)$ is the 
$\phi $-component of the Pointing vector, 
$\beta _n^2=\epsilon \omega _n^2/c^2-k_n^2$ is defined by the dispersion law of the $n$th mode, and $r$ is the radius 
of the cylindrical metal wire. Assuming $f(\omega )$ from (14) and $\beta _n\sim k_n$ the order of magnitude estimate of $W$ may be written as
\begin{equation}
W=zn^2(\frac{e^2}{\hbar c})\hbar \omega _n (\frac{\omega _p^4}{\omega 
_n^4})(\frac {\lambda }{r})\frac{\hbar ^2}{r^3m^2c^2},
\end{equation}
where $\lambda $ is the wavelength of $\omega _n$ light in vacuum. As may be 
expected, the nonlinear optical interaction of CSP modes grows inversely 
proportional to $r^4$, and achieves considerable strength in nanometer scale 
metallic nanowires and nanochannels even without addition of nonlinear 
dielectrics inside a nanochannel or around a nanowire. If we assume $\hbar 
\omega _n=1$eV, $\hbar \omega _p=7$eV, and $r=1$nm (disregarding nonzero skin 
depth) the potential energy of two unit chiral charges is of the order of $W\sim 
2\times 10^4z$eV/cm, so even interaction of single CSP quanta separated by 
submicrometer distances is considerable.

It is also important to mention that electron states with nonzero angular 
momenta in metal nanowires (such as gold, silver, copper, aluminum and many 
other metals which support surface plasmons, as well as carbon nanotubes with 
metallic conductance) should also behave like chiral charges described above. 
Their long-range current-loop-like interaction via exchange of $n=0$ CSPs must 
be taken into account while considering mesoscopic conductance of thin wires. 
The strength of such interaction should be much stronger than the interaction of 
CSPs with each other, since the potential energy of two electron current-loops 
is proportional to $f$ (and not to $f^2$ as in equation (24)). Long range 
attractive interaction between two electron current loops may have important 
consequences for superconductive properties of metal nanowires and carbon 
nanotubes, with considerable strength of such interaction indicative of 
potential high $T_c$ superconductivity.


\begin{references}

\bibitem{1} A. Imamoglu, H. Schmidt, G. Woods, and M. Deutsch, Phys.Rev.Lett. 
79, 1467 (1997). 

\bibitem{2} R.Y. Chiao and J. Boyce, Phys.Rev.A 60, 4114 (1999).

\bibitem{3} I.I. Smolyaninov, A.V. Zayats, A. Gungor, and C.C. Davis, 
Phys.Rev.Lett. 88, 187402 (2002).

\bibitem{4} I.I. Smolyaninov, A.V. Zayats, A. Stanishevsky, and C.C. Davis, 
cond-mat/0207450

\bibitem{5} L.D. Landau and E.M. Lifshitz, Electrodynamics of Continuous Media 
(Pergamon, New York, 1984).

\bibitem{6} T. Kaluza, Preus. Acad. Wiss. K 1, 966 (1921); O. Klein, Z.Phys. 37, 
895 (1926).

\bibitem{7} M.J. Duff, B.E.W. Nilsson, and C.N. Pope, Phys.Rep. 130, 1 (1986).

\bibitem{8} A. Chodos and S. Detweiler, Phys.Rev. D 21, 2167 (1980).

\bibitem{9} W. Schleich, M.O. Scully, in: G. Grynberg, R. Stora (Eds.), New 
Trends in Atomic Physics (North-Holland, Amsterdam, 1984) p.997.

\bibitem{10} D.F. Nelson, J.Appl.Phys. 86, 5348 (1999).

\bibitem{11} M. Born and K. Huang, Dynamical Theory of Crystal Lattices 
(Clarendon, Oxford, 1966), pp.336-338.

\bibitem{12} U. Schroter and A. Dereux, Phys.Rev.B 64, 125420 (2001).

\end{references}
\end{document}